\documentclass[prl,twocolumn,superscriptaddress,showpacs]{revtex4b5}
\usepackage{times,alg}
\usepackage[dvips]{graphicx,color}

\newcommand{\ket}[1]{|{#1}\rangle}
\newcommand{\bra}[1]{\langle{#1}|}

\newcommand{\slb}[2]{{#1}^{({#2})}}

\newcommand{\cE}{{\cal E}}

\newcommand{\mbo}{\mathbf{1}}
\newcommand{\ignore}[1]{}

\newcommand{\minpolarization}{48\%}
\newcommand{\maxpolarization}{87\%}
\newcommand{\avpolarization}{67\%}

\newcommand{\goalbcfe}{.75}
\newcommand{\goalbcerror}{.02}
\newcommand{\noisegoal}{8\%}
\newcommand{\solvent}{acetone}

\newcommand{\phead}[1]{\par\noindent{\bf #1}}

\begin{document}

\title{Implementation of the Five Qubit Error Correction Benchmark}
\author{E. Knill}
\email{knill@lanl.gov}
\author{R. Laflamme}
\email{laflamme@lanl.gov}
\author{R. Martinez}
\author{C. Negrevergne}
\affiliation{Los Alamos National Laboratory, MS B265, Los Alamos, New Mexico 87545}
\date{\today}

\begin{abstract}
The smallest quantum code that can correct all one-qubit errors is
based on five qubits. We experimentally implemented the encoding,
decoding and error-correction quantum networks using nuclear magnetic
resonance on a five spin subsystem of labeled crotonic acid. The
ability to correct each error was verified by tomography of the
process. The use of error-correction for benchmarking quantum networks
is discussed, and we infer that the fidelity achieved in our
experiment is sufficient for preserving entanglement.
\end{abstract}

\maketitle

Robust quantum computation requires that information be encoded to
enable removal of errors unavoidably introduced by
noise~\cite{shor:qc1995b,steane:qc1995a,knill:qc1995e,shor:qc1996a,aharonov:qc1996a,kitaev:qc1997a,knill:qc1998a,preskill:qc1998a}.
Thus, every currently envisaged scalable quantum computer has
encoding, decoding and error-correction procedures among its most
frequently used subroutines. It is therefore critical to verify the
ability to implement these procedures with sufficient fidelity. 
The experimental fidelities achieved serve as useful benchmarks to
compare different device technologies and to determine to what extent
scalability can be claimed.

Liquid state nuclear magnetic resonance (NMR) is currently the only
technology that can be used to investigate the dynamics of more than
four qubits~\cite{cory:qc1997a,chuang:qc1997a}. Although it is not
practical to apply it to more than about ten
qubits~\cite{warren:qc1997a}, it can be used to investigate the
behavior of quantum networks on representative physical systems to
learn more about and begin to solve the problems that will be
encountered in future, more scalable device technologies.  In this
Letter, we describe an experimental implementation using NMR of a
procedure for benchmarking the one-error correcting five-qubit code.
This is the shortest code that can protect against depolarizing
one-qubit errors~\cite{bennett:qc1996a,laflamme:qc1996a}.  The
experiment is one of the most complex quantum computations implemented
so far and the first to examine the behavior of a quantum
error-correcting code which protects against all one-qubit quantum
noise. We discuss the principles underlying error-correction
benchmarks and offer a sequence of specific challenges to be met by
this and future experiments. The ultimate challenge is to demonstrate
the ability to reduce the destructive effects of independent
depolarizing errors. This seems to be out of reach of liquid state NMR
experiments. Our experiment shows an average polarization preservation
of $\avpolarization$ corresponding to an entanglement fidelity of
$\goalbcfe$. This easily achieves the goal of demonstrating
the preservation of entanglement in principle.

\phead{The five-qubit code.}  A quantum error-correcting code for
encoding a qubit is a two-dimensional subspace of the state space of a
quantum system. In the case of interest, the quantum system consists
of five qubits. The code, $C_5$, can be specified as one of the 16
two-dimensional joint eigenspaces of the four commuting operators
\begin{eqnarray}
\slb{\sigma}{2}_z 
\slb{\sigma}{3}_y
\slb{\sigma}{4}_y
\slb{\sigma}{5}_x
&,&
\slb{\sigma}{1}_z
\slb{\sigma}{2}_y
\slb{\sigma}{3}_y
\slb{\sigma}{4}_x
,\nonumber\\
\slb{\sigma}{2}_y
\slb{\sigma}{3}_z
\slb{\sigma}{4}_z
\slb{\sigma}{5}_z
&,&
\slb{\sigma}{1}_x
\slb{\sigma}{2}_z
\slb{\sigma}{3}_x
\slb{\sigma}{4}_z
.
\label{eqn:q5stab}
\end{eqnarray}
Here, $\slb{\sigma}{k}_x,\slb{\sigma}{k}_y,\slb{\sigma}{k}_z$ are the
Pauli spin operators acting on qubit $k$.  This is an instance of a
stabilizer code~\cite{gottesman:qc1996a,calderbank:qc1996a}, and the
fact that it can be used to correct any one-qubit error is due to the
property that every product of one or two Pauli operators acting on
different qubits anticommutes with at least one of the operators in
Eq. (\ref{eqn:q5stab}).

\begin{figure}[p]
\includegraphics[width=3.2in]{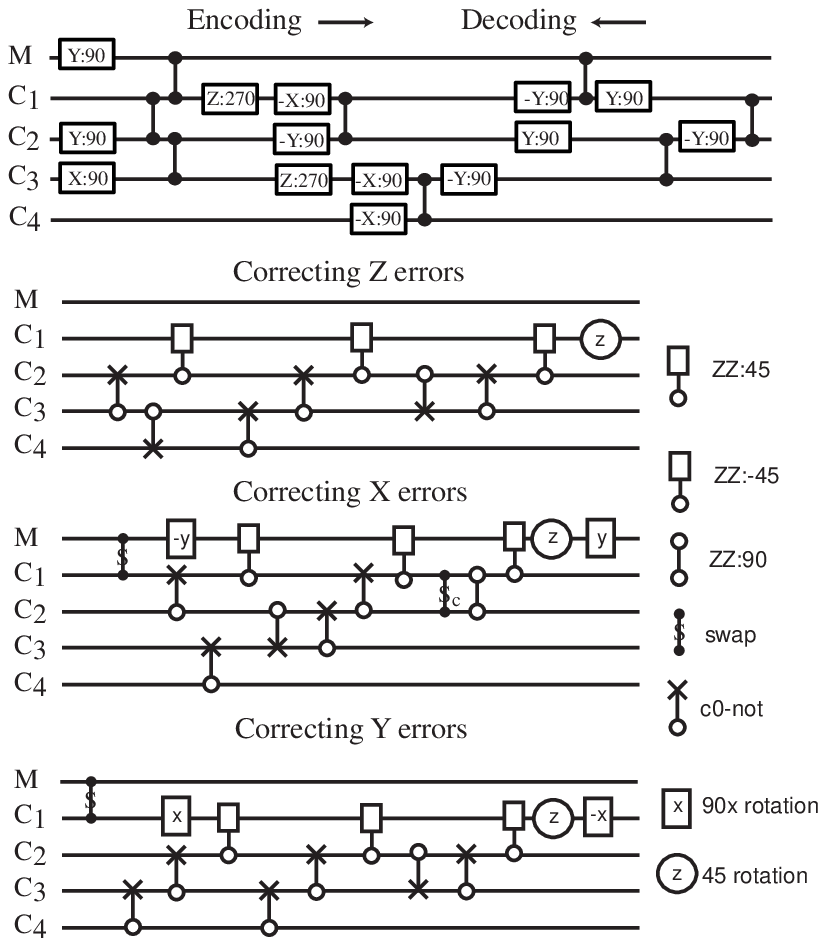}
\caption{Networks for the 5-qubit code.
Top: The encoding network using $90^\circ$ rotations.
Except for refocusings required to eliminate unwanted couplings,
these are directly implementable with pulses. The decoding
network is the inverse of the encoding network.
Bottom: The three steps of the error-correction procedure, which
implements a rotation on C1 conditional on the syndrome state.
The controlled operations can be translated to sequences
of $90^\circ$ rotations using standard quantum network
methods~\cite{somaroo:qc1998a}.
}
\label{fig:networks}
\end{figure}

A typical application of a quantum error-correcting code is to protect
a qubit's state in a noisy quantum memory. As implemented in our
experiment, the procedure begins with qubit $2$ containing the state
to be protected and \emph{syndrome} qubits $1,3,4,5$ in the initial
state $\ket{1}$. A unitary \emph{encoding} transformation is applied
to map the two-dimensional input state space to the subspace of the
code $C_5$. In the application, the five qubits are then stored
in the noisy memory. In our experiment, we explicitly applied one of the
correctable error. The information is retrieved by first applying the
inverse of the encoding transformation to \emph{decode} the state. The
properties of the code guarantee that if the error was a Pauli
operator acting on one of the qubits, which one occurred is reflected
in the state of the syndrome qubits. To complete the process,
conditional on the syndrome qubits' state, it is necessary to
\emph{correct} the state of qubit $2$ by applying a Pauli
operator. Quantum networks for encoding, decoding and error-correction
are shown in Fig.\ref{fig:networks}.

\phead{Benchmarking quantum codes.}

The purpose of a benchmark is to compare the performance of different
devices on the same task. Since quantum codes will be used to maintain
information in future quantum computing devices, they are excellent
candidates for benchmarking the reliability of proposed quantum
processors. A basic quantum code benchmark consists of a sequence of
steps that implement encoding, evolution, decoding and
error-correction networks. In the simplest cases, the steps are
applied to one qubit's state using several ancilla qubits for the
intermediate steps. An experimental implementation measures the
reliability (see below) with which the qubit's state is processed. It
is necessary to include a means for verifying that a code with the
desired properties was indeed implemented. For error-correcting and
for stabilizer codes, this can be done by inserting $180^\circ$ pulses
applied to individual qubits in the evolution step and observing their
effect on the output. In this case, the verification relies on the
assumption that such pulses can be reliably implemented, a property
that needs to be independently verified.

To allow for unbiased comparison of devices, the reliability
measurement and the verification steps of the benchmark need to be
standardized. There are many different ways of quantifying
reliability. The best known such measure, \emph{fidelity}, is based on
the geometry of the state space.  If the input state is $\ket{\psi}$
and the output density matrix is $\rho$, then the fidelity of the
output is given by $F(\ket{\psi},\rho)=\bra{\psi}\rho\ket{\psi}$. This
can be seen to be the probability of measuring $\ket{\psi}$ in a
measurement which distinguishes this state from the orthogonal states.
In our case we are interested in an arbitrary state of one qubit.  One
quantity of interest would be the worst case pure state fidelity,
which minimizes the fidelity of the implemented benchmark over pure
input states. However, an easier to use quantity is the \emph{entanglement
fidelity} $F_e$~\cite{schumacher:qc1996a}, the fidelity with
which a Bell state on the qubit and a perfect reference qubit is
preserved.  Entanglement fidelity does not depend on the choice of
Bell state and has the property that $F_e=1$ if and only if the
process perfectly preserves every input state.  To avoid
experimentally implementing the reference qubit, $F_e$ can be
determined from the fidelities of pure states.  Define
$\ket{\pm}=(\ket{0}\pm\ket{1})/\sqrt{2}$ and $\ket{\pm i}=(\ket{0}\pm
i\ket{1})/\sqrt{2}$ (the eigenstates of $\sigma_x$ and $\sigma_y$,
respectively). Let $F_{s}$ be the fidelity of the process
for input $\ket{s}$. Then
\begin{equation}
F_e = (F_0+F_1+F_+ + F_- + F_{+i} + F_{-i})/4-1/2.
\end{equation}
We advocate the use of entanglement fidelity as the standard
reliability measure to be given when describing the results
of a quantum benchmark involving processing of a quantum bit.

The standard verification procedure for a quantum code benchmark needs
to be such that sufficiently high fidelity cannot be achieved without
having implemented a code with the desired properties.  For codes
defined as the common eigenspace of a commuting set of products of
Pauli operators (stabilizer codes), it is in principle enough to verify
that applying a product $P$ of Pauli operators during the evolution
has the expected effect on the output, namely that it differs from the
input by the application of a Pauli operator $\sigma(P)$ determined by
the code and the applied product. A single fidelity measure may be
obtained by applying $\sigma(P)$ to the output and by averaging the
resulting entanglement fidelities over all $P$.  To make this
procedure experimentally feasible, one may randomize the choice of $P$
and use statistical methods to estimate the desired average.

For benchmarks involving a quantum error-correcting code, the emphasis
is on having corrected the set of errors $\cE$ for which it was
designed, and verification involves applying the errors in $\cE$
during the evolution and observing the extent to which they are indeed
corrected. Ideally, the errors occur naturally in the course of
evolution, and one would like to see that information is preserved
better by encoding it. In order to investigate the code in a
controlled way, it is easier to apply different errors explicitly and
observe the effect on the reliability of the process.
The experiment described here involves measuring the entanglement
fidelity for each of the one-qubit Pauli operators applied during the
evolution.

\begin{figure}[p]
\includegraphics[width=3.2in]{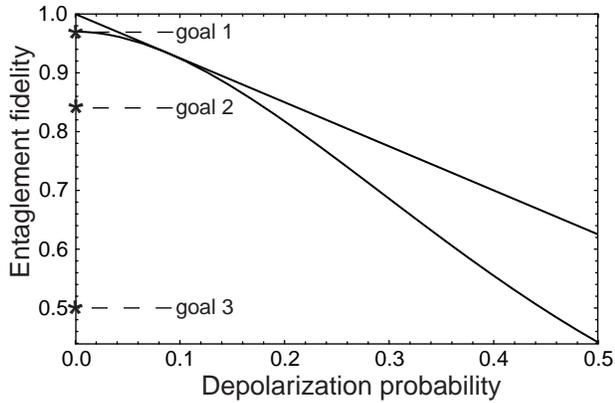}
\caption{Entanglement fidelities for independent depolarization.  The
fidelities for an unencoded qubit (straight line) and a qubit encoded
with $C_5$ are shown as a function of the depolarization probability.
The implementation of the code is assumed to have an additional error
that is syndrome independent with depolarization probability $.97$.
The first three fidelity benchmarking goals are indicated.}
\label{fig:goals}
\end{figure}

When implementing a benchmark, it is useful to have some goals in
mind. Each goal implies the demonstration of a non-trivial
result. For benchmarks involving codes designed to correct independent
errors on qubits, we offer a sequence of four such goals depending on
how well the implementation succeeds at protecting against various
error models.  Most involve comparing the entanglement fidelity for
two situations involving a specific error process $\cE_i$. In the
first, the information is stored in any one of the qubits, giving an
optimum $F_{e,1}(\cE_i)$. In the second, the information is stored by
using the implemented code, giving an experimentally determined
$F_{e,C}(\cE_i)$. Numerical goals for $F_{e,C}(\cE_i)$ are given under
the assumption that the implementation induces depolarizing noise.
The goals are:
1.  Improvement where $\cE_1$ is depolarization of each qubit with
some probability $p$.  For $C_5$ under the assumption that the
error behavior is identically depolarizing (quantified by $F_{e,C_5}$)
for each possible Pauli-product error during the evolution,
this requires $F_{e,C_5} > 0.97$, giving an improvement when
$p=0.08713$. See Fig.~\ref{fig:goals}.
\ignore{
unenc = [];
enc = [];
ps = [];
fe = .97;
for k=(1:21);
 p = (k-1)/40;
 ps(k) = p;
 enc(k) = (1-3/4*p);
 unenc(k) = (4*fe*(-1 + p)^3*(-2 - 6*p + 3*p^2 ) + ...
    p^2 *(15 - 25*p + 15*p^2  - 3* p^3 ))/8;
end;
fig = figure(2);
hold off;
plot(ps,unenc,'k');
hold on;
plot(ps,enc,'k');
}
2. Improvement where $\cE_2$ is the process that first randomly
chooses a qubit and then depolarizes it.  For our code this requires
$F_{e,C_5}(\cE_2) > .85$.
3. Preservation of some entanglement for $\cE_2$. This requires
$F_{e,C}(\cE_2) > .5$~\cite{bennett:qc1996a}.
4. Improvement in the presence of the \emph{demonic} error process
$\cE_4$ that, knowing the method for storing the qubit, chooses the
worst possible one-qubit depolarizing error and applies it.  In this
case we need $F_{e,C}(\cE_4) > .25$.
One of
the ultimate goals might be to demonstrate that the code can be
implemented sufficiently well to permit preservation of
information by means of concatenation. 

\begin{figure}[p]
\includegraphics{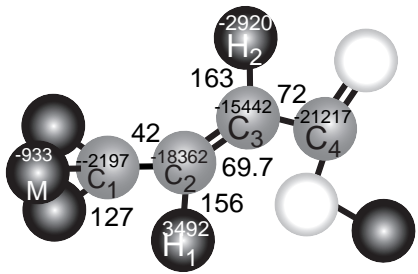}
\caption{Trans-crotonic acid. The chemical shifts and nearest neighbor
couplings are shown.}
\label{fig:molecule}
\end{figure}

\phead{Experimental implementation.}  We used $^{13}$C labeled
trans-crotonic acid (Fig.~\ref{fig:molecule}) synthesized as described
in~\cite{knill:qc1999a}, but with deuterated {\solvent} as a
solvent. A standard 500 MHz NMR spectrometer (DRX-500 Bruker
Instruments) with a triple resonance probe was used to run the
experiments. The five spin $1/2$ systems used for the code are the
methyl group (M), C1, C2, C3 and C4. The methods
of~\cite{knill:qc1999a} were used to prepare the methyl group as an
effective spin $1/2$ system and to initialize the labeled pseudo-pure
state $\mbo\sigma_z\mbo\mbo\mbo\mbo\mbo$ on all active nuclei.  Here,
$\mbo = \ket{1}\bra{1}$ and the last two nuclei are H1 and H2. The
selection method was based on gradients, and the pseudo-pure state was
subjected to a ``crusher'' gradient. To absolutely guarantee the
pseudo-pure state, more randomization is required
(see~\cite{knill:qc1999a}) but we did not implement this. H1 and H2
were not used and were only affected by some hard pulse refocusings on the
protons. The state of H1 and H2 (up or down) induces an effective
frequency shift on the other nuclei depending on the coupling
constants and was compensated for in phase calculations.  To greatly
reduce the effect of radio frequency (RF) inhomogeneity, we used the
nutation based selection scheme of~\cite{knill:qc1999a}, applied to
both the proton and the carbon transmitters.  The quantum networks of
Fig.~\ref{fig:networks} were directly translated to pulse sequences,
again using the methods described in~\cite{knill:qc1999a}. The only
significant use of manual intervention was to place the refocusing
pulses.  The evolution period between encoding and decoding was
carefully isolated from both the preceding and the following pulses:
It implemented the identity unitary operator, or one of the one qubit
$180^\circ$ rotations by refocusing the molecule's internal Hamiltonian
and applying an extra inversion or by shifting the phase by
$180^\circ$. The qubit's output state appeared on C1 at the end of the
experiment. The peak group associated with C1 was observed in each
experiment. Spectra were analyzed by comparing the spectrum of the
pseudo-pure state $\mbo\sigma_x\mbo\mbo\mbo\mbo\mbo$ to the output,
using the knowledge of the peak positions and shapes to compute
relative intensities and phases. \emph{No} phase adjustment was made
after phasing the pseudo-pure state spectrum. This was possible since
the relative phase is precomputed by the pulse compiler and
integrated into the acquisition.

We performed one experiment for each of the $16$ possible evolutions
with one-qubit or no Pauli error, each of the three initial states
$\sigma_x$, $\sigma_y$ or $\sigma_z$ on C1, and each of three
observations (no pulse, $90^\circ$ $X$ pulse, or $90^\circ$ $Y$ pulse)
on C1.  This resulted in a total of 112 experiments, each of which was
repeated sufficiently often to get better than $\noisegoal$ error in
the inferred state of C1.  For each evolution $E$ and input
$\sigma_u$, we determined the amount of signal $P(e,u)$ in the correct
direction in the output relative to the input signal. This requires
``tracing out'' the other spins, which was done by adding the
intensities of each peak in the C1 spectrum that is associated
with the $\mbo\mbo$ state on $H1$ and $H2$. (The spectrum of C1
resolves all its couplings.)  Thus, except for noise, $-1\leq
P(E,u)\leq 1$. Under the assumption that input $\mbo \openone
\mbo\mbo\mbo\mbo\mbo$ results in no observable signal, the
entanglement fidelity for a given evolution $e$ is given by $F_e(E) =
(P(E,x)+P(E,y)+P(E,z)+1)/4$.  We did not verify the assumption in this
experiment, but note that it has been verified in related
experiments~\cite{cory:qc1998a}, and could be enforced by modifying
the process with random pairs of canceling $180^\circ$ pulses before and after
the implemented pulse sequence. This would also enforce the depolarizing
noise model for the implementation while preserving the observed
polarization.

\begin{figure}[p]
\includegraphics{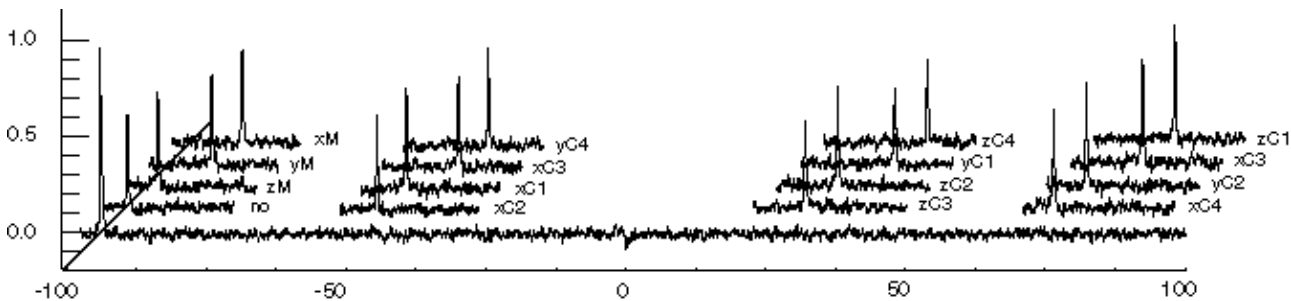}
\caption{Experimental input and output spectra.  The reference
spectrum for the pseudo-pure input is at the bottom, and partial
spectra for each one-qubit error are shown above it using the same
scale. The labels indicate which error on which nucleus was applied.
One peak is observed for each possible error input. Its position
corresponds to the error syndrome. Its phase reflects the error
correction procedure and corresponds to the input state up to a small
error. Signal in the wrong locations or phase was consistently small
and comparable to the estimated noise.  }
\label{fig:spectra}
\end{figure}

\begin{figure}[p]
\includegraphics[width=3.2in]{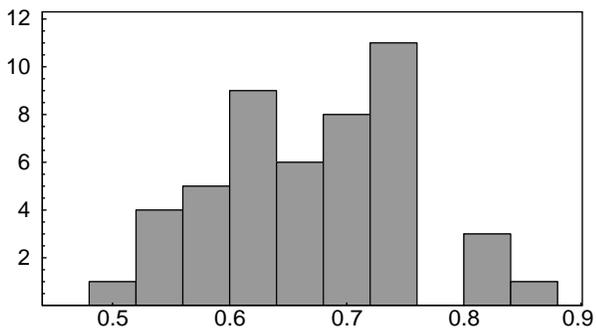}
\caption{Distribution of relative polarizations. There are a
total of $48$ polarization measurements. Each bar represents
the number of measurements with relative polarization
in the bar's interval. The distribution strongly suggests
some syndrome-dependent effects on the implementation error.
}
\label{fig:pol_dist}
\end{figure}

\phead{Results.}  Typical spectra compared to the spectrum of the
input pseudo-pure state are shown in Fig.~\ref{fig:spectra}.  The
relative polarization after the error-correction procedure in the
correct output state varies between $\minpolarization$ and
$\maxpolarization$. The distribution is shown in
Fig.~\ref{fig:pol_dist}. The inferred entanglement
fidelity for goals $2$ and $3$ is $F_{e,C}(\cE_2)=\goalbcfe$,
with an estimated error of less than $\goalbcerror$.
Thus we clearly met goal $3$.

The reduction in polarization is due to thermal relaxation,
incompletely refocused couplings (part of the pulse compiler's
optimization trade-offs), pulse errors due to non-ideal implementation
of $180^\circ$ and $90^\circ$ pulses, and RF inhomogeneity (less than $2\%$
after our selection procedure).  Most of the error is explained by the
known relaxation times, suggesting that this is what limits
the fidelities that can be attained using liquid state NMR. We note
that the estimated phase relaxation times of well over a second are
high when compared to those of nuclei in other molecules used in NMR
quantum information processing experiments to date.

\phead{Discussion.}  Benchmarking quantum devices for quantum
information processing is crucial both for comparing different device
technologies and for determining how much control over a device is
achievable and how to best achieve it. Given the need for and
difficulty of achieving robust quantum control, we advocate the use of
quantum coding benchmarks to determine the fidelity of the
implementation of standard, verifiable processes.  Unlike the
experimentally implemented versions (up to 5 qubits) of the popular
quantum searching and order-finding
algorithms~\cite{jones:qc1998a,jones:qc1998b,chuang:qc1998a,chuang:qc1998c,marx:qc1999a,vandersypen:qc2000a},
quantum codes offer a rich source of complex and verifiable quantum
procedures required in currently envisioned quantum computer
architectures. Liquid state NMR has been used to implement several
interesting, small quantum codes~\cite{cory:qc1998a,leung:qc1999b},
and a seven qubit cat-state benchmark~\cite{knill:qc1999a}.  In this
letter, we have given specific goals for benchmarks involving
error-correction and, for the first time, implemented a one-qubit
error-correcting quantum code. The fidelity achieved is well above
what is needed to demonstrate preservation of entanglement in the
presence of a non-trivial error-model. It is unlikely that the goal of
showing improvement in error-correcting independent depolarizing
errors is achievable with liquid state NMR.  A device that achieves
this challenging goal will be well on the way toward realizing
robustly scalable quantum computation.

\phead{Acknowledgments.} Supported by the Department of Energy
(contract W-7405-ENG-36) and by the NSA. We thank the Stable Isotope
Resource for providing equipment and support. 
We thank Ryszard Michalczyk, Brian MacDonald, Cliff Unkefer and David Cory
for their help.

\bibliography{journalDefs,qc}

\pagebreak

\end{document}